\documentclass[12pt]{article}
\usepackage{amsmath,amssymb,epsfig}
\usepackage{cancel}
\usepackage{color}
\usepackage{bm}
\usepackage{braket}

\makeatletter

\@addtoreset{equation}{section}
\makeatother

\begin{document}

\title{\vbox{
\baselineskip 14pt
\hfill 
\hbox{\normalsize EPHOU-17-008}\\ 
\hfill \hbox{\normalsize KUNS-2678} \\
\hfill \hbox{\normalsize MISC-2017-05} 
\vskip 1.7cm
\bf Poly-instanton axion inflation \vskip 0.5cm
}}
\author{
Tatsuo~Kobayashi$^{1}$, \
Shohei~Uemura$^{2}$, \ and \
Junji Yamamoto$^{3}$
\\*[20pt]
\\
{\it \normalsize 
${}^{1}$Department of Physics, Hokkaido University, 
Sapporo, 060-0810 Japan}
\\
{\it \normalsize 
${}^{2}$Maskawa Institute for Science and Culture, Kyoto Sangyo University, Kyoto 603-8555, Japan}
\\
{\it \normalsize 
${}^{3}$Department of Physics, Kyoto University, 
Kyoto 606-8502, Japan}
\\*[50pt]}

\date{
\centerline{\small \bf Abstract}
\begin{minipage}{0.9\linewidth}
\medskip 
\medskip 
\small
We investigate the axion inflation model derived by poly-instanton effects in type II superstring theories.
Poly-instanton effects are instanton effects corrected by another instanton and it can generate the modulus-axion potential with the 
double exponential function.
Although the axion has a period of small value, this potential can have a flat region because its derivatives  are exponentially suppressed by non-perturbative effects.
From the view point of the cosmic inflation, such potential is interesting.
In this paper, we numerically study the possibilities for realizing the cosmic inflation.
We also study their spectral index and other cosmological observables, numerically.
\end{minipage}
}

\newpage

\begin{titlepage}
\maketitle
\thispagestyle{empty}
\clearpage
\end{titlepage}

\renewcommand{\thefootnote}{\arabic{footnote}}

\section{Introduction}

The cosmic inflation is the important stage in the history of the Universe solving the problems 
of the standard Big Bang cosmology such as the flatness problem and horizon problem, and deriving 
the observed fluctuation of the cosmic microwave background.
From the theoretical viewpoint, 
what is the field deriving the inflation, i.e. inflaton, is 
one of key questions to understand the Universe from underlying theory such as 
superstring theory.
It should have sufficiently flat potential.
Indeed, various studies have been done  within the framework of  superstring theory \cite{Baumann:2014nda}.

Axions as well as moduli are good candidates for the inflaton, 
because axions have continuous shift symmetries.
Non-perturbative effects break  continuous shift symmetries to discrete ones, 
and the potential of the axion can be generated.
However, such a potential would be controlled well by discrete shift symmetries.
Thus, non-perturbative effects are quite important.
Non-trivial backgrounds can also break continuous shift symmetries, 
generating the axion potential.

The natural inflation \cite{Freese:1990rb} and the axion monodromy inflation \cite{Silverstein:2008sg}\footnote{
See also \cite{Kaloper:2008fb}.} are 
of famous axion inflation models.
The simple natural inflation requires a large decay constant, although 
the typical decay constant would be smaller than the string and Planck scales.
A large decay constant can be realized effectively, e.g. by
the alignment mechanism \cite{Kim:2004rp} and  loop effects \cite{Abe:2014pwa,Abe:2014xja}.
The simple natural inflation may lead to disfavored predictions on 
observables, but modulation terms would change significantly predictions 
on observables \cite{Abe:2014xja,Czerny:2014wza,Kobayashi:2016vcx}.
Correction terms in monodromy inflation are also important and they 
change significantly predictions on observables \cite{Kobayashi:2016vcx, Kobayashi:2014ooa}.

String world-sheet instaton corrections are stringy-calculable effects.
The world-sheet instanton effects can be 
controlled by modular symmetries, which are characteristic symmetries in 
string theories.
Then, certain modular functions appear as the potential of moduli and axion.
The inflation scenario with such modular function terms has been also studied \cite{Abe:2014xja,Kobayashi:2016mzg,Higaki:2016ydn}.
Moreover, the mixture of polynomial functions and sinusoidal functions 
was derived by quantum corrections and studied to realize inflation models \cite{Kobayashi:2015aaa,Parameswaran:2016qqq,Kadota:2016jlw}.

Furthermore, one can compute couplings of moduli and axions to gauge bosons 
and quarks and leptons in four-dimensional (4D) low energy effective field theory.
Thus, we can explicitly study the thermal history after the axion inflation.
For example, we can analyze explicitly 
the reheating processes after the axion inflation.
The axion couplings to the gauge bosons and quarks and leptons may be rather weak, 
and the reheating temperature may be rather low \cite{Kobayashi:2016vcx, Kobayashi:2015aaa,Kadota:2016jlw}.
The axions would also appear in phases of couplings among quarks, leptons and the Higgs fields.
After the inflation, the axion would oscillate, and then coupling phases would vary.
Such an oscillation would have important effects on the history of the Universe.
For example, the baryon asymmetry may be generated by such oscillation \cite{Akita:2017ecc}.

As mentioned above, various types of the axion inflation models have been studied and 
several aspects have been discussed after the inflation.
It would be interesting to study various possibilities for the axion inflation, 
and their several aspects.
Recently, a new type of non-perturbative effects was studied, that is, the so-called poly-instanton \cite{Blumenhagen:2008ji, Blumenhagen:2012kz}.
That leads to the exponential function of exponential function as the 
moduli potential term.
It would be interesting to study the possibility for the axion inflation 
by use such a new type of the potential.
Indeed, the moduli inflation has been studied by use of such a new potential term \cite{Cicoli:2011ct}.\footnote{
The moduli stabilization has been also studied \cite{Cicoli:2011yy,Blumenhagen:2012ue}.}
The purpose of this paper is to study the possibilities  for the axion inflation 
by use the new type of the axion potential.

This paper is organized as follows.
In section 2, we show which type of axion potential is generated by 
poly-instanton effects.
In section 3,  we study numerically the poly-instanton axion inflation.
Section 4 is conclusion.
In Appendix A, we discuss some examples, that only the axion remains light, 
while the other moduli are stabilized by heavy masses.

\section{Poly-instanton potential}

The 4D low energy effective field theory of superstring theory is 
described by supergravity theory, and it includes 
the dilaton $S$, the K\"ahler moduli $T_i$ and complex structure moduli $U_m$.
Here, we focus on type IIB superstring theory, although our scenario 
would be applicable to other superstring theories.
Hereafter, we use the unit that $M_{Pl} =1$, where $M_{Pl}$ denotes 
the reduced Planck scale, i.e. $M_{Pl} = 2.4 \times 10^{18}$~GeV.

The superpotential $W$ can be written by a combination of perturbative and 
non-perturbative terms,
\begin{equation}
W=W_{p}+ W_{np}.
\end{equation}
The gaugino condensations and D-brane instanton effects are well-known computable non-perturbative effects 
in type II superstring theory.
The superpotential terms induced by  D-brane instantons can be written as \cite{Cvetic:2007ku, Blumenhagen:2009qh},
\begin{equation}\label{eq:W-inst}
W_{np} = \sum_i A_i(U_m) e^{-a_i {\cal S}_i}.
\end{equation}
Here, ${\cal S}_i$ denotes the action of the D-brane instantons, and it can be obtained as 
${\cal S}_i=\int_{\Gamma_i} dV = M S + m_{i,j} T_j$, where 
$\Gamma_i$ represents a p-cycle wrapped by the D-brane instanton and $m_{i,j} T_j$ represent the K\"ahler moduli  corresponding to the volume of $\Gamma_i$.
Moduli $T_i$ are written as,
\begin{equation}
T_i = \tau_i + i \phi_i,
\end{equation}
and the imaginary parts $\phi_i$ represent axions.
In addition, $a_i$ is a constant parameter and its value is typically  2$\pi$ for D-brane instantons.
In general, $A_i(U_m)$ is a function of the complex structure moduli $U_m$.
Here, we assume that all of the complex structure moduli $U_m$ and the dilaton $S$ are stabilized 
by three-form flux background \cite{Giddings:2001yu, Kachru:2003aw}.

On the other hand, the gaugino condensation in the hidden sector $G_a$  would generate 
the following term:
\begin{equation}
W_{np} = A_ae^{-24 \pi^2 f_a/ b_a},
\end{equation}
where $f_a$ and $b_a$ are gauge kinetic function and beta-function coefficient of $G_a$, respectively.
The real part of the gauge kinetic function provides with the gauge coupling $g_a$, i.e. $1/g_a^2 = Re(f_a)$. 
Including quantum corrections, the gauge kinetic function can be written by 
\begin{equation}
f_a = f_{tree} + f_{1-loop} + f_{np}.
\end{equation}
Here, the tree-level gauge kinetic function $f_{tree}$ is a linear combination of $S$ and $T_i$ 
 corresponding to the volume of the cycle, where the hidden sector $G_a$ wraps. 
The 1-loop correction $f_{1-loop}$ is a function of complex structure moduli $U_m$ and independent of K\"ahler moduli $T_i$.
The non-perturbative correction, $f_{np}$ is given by
\begin{equation}
f_{np}=\sum_j B_j(U_m) e^{-b_j T_j},
\end{equation}
and such a correction can be induced by D-brane instantons having one complex position moduli on the compact space.
Thus, in general, we can write non-perturbative terms in the superpotential,
\begin{equation}
W_{np} =  \sum_a A_a \exp [-b_a (\alpha_{a,i} T_i +\sum_{j}  B_{a, j} e^{-b_{j,k} T_k})],
\end{equation}
where we have redefined coefficients including VEVs of $S$ and $U_m$, because 
we have assumed that the dilaton and complex structure moduli are stabilized.
A similar non-perturbative superpotential term would appear through the D-brane instanton terms 
(\ref{eq:W-inst}) with corrections.
Since it is an instanton correction corrected by other instanton effects, it is called poly-instanton \cite{Blumenhagen:2008ji, Blumenhagen:2012kz}.
It can generate potential for K\"ahler moduli and stabilize them \cite{Cicoli:2011ct, Cicoli:2011yy, Blumenhagen:2012ue}.

The $F$-term scalar potential is given by,
\begin{equation}\label{eq:V}
V_F = e^K [K^{T_{i} \bar T_{j}} D_{T_{i}} W D_{\bar{T_{j}}} \bar{W}  -3|W|^2], 
\end{equation}
where $K$ denotes the K\"ahler potential and $D_{T_i}W = (\partial_{T_i}K)W + \partial_{T_i} W$.
The K\"ahler potential  is given by $K=-2 \ln \mathcal{V}$ where $\mathcal{V}$ is the volume of the compact space, 
and the  K\"ahler potential is a function of $T_i + \bar T_i$, i.e. $K(T_i + \bar T_i)$.
In addition, $K^{T_{i} \bar{T_j}}$ represents the inverse of $K_{T_{i} \bar{T_j}} = \partial_{T_i} \partial_{\bar{T_j}}K$.

In this paper, we focus on the axion potential generated by the poly-instanton.
Suppose that there is a modulus field $T_0$, which appears only in poly-instanton superpotential, but dose not appear at the tree level and single instanton potential, i.e.,
\begin{equation}
W=W(T_i)|_{i\neq 0}+ A_0 \exp [-a ( T_i + C e^{-a_{0} T_{0}})]|_{i\neq 0}.
\end{equation}
When $A_0=0$, $\phi_0$ remains massless, since $\phi_{0}$ does not appear in the $F$-term scalar potential.
However, all the other moduli including $\tau_0$  appear in the potential. 
Here, we assume that these moduli including $\tau_0$ are stabilized with heavy masses.
We show some models realizing this assumption in Appendix \ref{App:example}.
Their masses would be estimated by $W(\braket{T_i})|_{i\neq 0}$.
Thus, when $A_0 \ll W(\braket{T_i})|_{i\neq 0}$, the mass of $\phi_{0}$ is much smaller than those of the other moduli fields.
Therefore, we integrate all the other fields out, and obtain the superpotential,
\begin{equation}\label{eq:poly-super}
W=W_0' + A_0' \exp [ -a C e^{-a_{0} T_{0}}],
\end{equation}
where $W_0'= W(\braket{T_i})|_{i\neq 0}$ and $A_0'=A_0 \exp[ -a \braket{T_i}]$.

%
%

%

The $F$-term scalar potential derived from (\ref{eq:poly-super}) has a very specific form.
The second term, $-3 e^{K} |W|^2$, is calculated as,
\begin{equation}
-3e^K|W|^2 = -3 e^K(|W_0'|^2 + \bar W_0' A_0' \exp [ -a C e^{-a_{0} T_{0}}] + h.c. ) + \mathcal{O}(A'^2_0),
\end{equation}
where we have used the approximation, $|W_0'| \gg |A_0'|$.
Similarly, we approximate 
\begin{equation}
\begin{split}
e^KK^{T_{0} \bar T_{0}} |D_{T_{0}} W|^2 \approx& ~e^KK^{T_{0} \bar T_{0}} \{ K_{T_{0}}^2(|W'_0|^2 + \bar W'_0 A'_0 \exp [  -a C e^{-a_{0}T_{0}}] + h.c. )\\
& + K_{T_{0}} \bar W'_0 A'_0a_{0} a Ce^{-a_{0} T_{0}} \exp [ -a C e^{-a_{0}T_{0}}] + h.c. \} +\mathcal{O}(A'^2_0).
\end{split}
\end{equation}
Then, we obtain the $F$-term scalar potential for the axion generated by a poly-instanton,
\begin{equation}\label{eq:V_axion}
\begin{split}
V \approx & 
  ~2e^K|\bar W_0' A'_0 | \exp [ -\delta \cos(a_{0} \phi_{0})] \{ (K^{T_{0} \bar T_{0}}K_{T_{0}}^2 - 3)  \cos[ \delta \sin (a_{0} \phi_{0}) +\theta ]\\
& + K^{T_{0} \bar T_{0}} K_{T_0} a_{0} \delta \cos[ -a_{0} \phi_{0} + \delta \sin (a_{0} \phi_{0})+\theta]\} + ({\rm const}) ,
\end{split}
\end{equation}
where $\delta= a C\exp[-a_{0} \braket{\tau_{0}}]$, and 
$\theta$ is the phase of the prefactors.
We omitted the terms of  ${\cal O}(A'^2_0)$.

In the next section, we study the cosmological inflation using the potential (\ref{eq:V_axion}), 
which is rather complicated.
We estimate that $K^{T_{0} \bar T_{0}}, K_{T_{0}}^2 ={\cal O}(1)$.
Here, we treat $\delta$ and $a_0$ as free parameters from the phenomenological viewpoint. 
When $a_0 \delta \gg 1$, the second term in the curly bracket is dominant.
Then, the potential for the canonically normalized axion $\phi=\sqrt{2 K_{T_0 \bar{T}_0} } \phi_0$  
is approximated to
\begin{equation}\label{eq:Vlarge_delta}
V = A (\,e^{-\delta \cos a \phi} \cos[\delta \sin a \phi- a \phi +\theta] +V_0\,). 
\end{equation}
Here, we redefined coefficients as,
$a= a_0/\sqrt{2 K_{T_0 \bar{T_0}}}$, and $A= 2 e^K\bar W_0' A'_0 K^{T_{0} \bar T_{0}}  a_{0} \delta$.
We also added the uplifting term $V_0$ to set the potential (local) minimum to be almost zero.
\footnote{
Such a uplifting term can be realized by supersymmetry breaking D-brane configuration \cite{Kachru:2003aw}
or spontaneous supersymmetry breaking, i.e., the so-called F-term uplifting  \cite{Lebedev:2006qq}.}
To cancel the former term, $V_0$ is of the order of $e^{-\delta_{min}}$, where $\delta_{min} = \delta \cos a \phi_{min}$ with 
the value $\phi_{min}$ at the potential (local) minimum.
In the next section, we will concentrate on the axion inflation with the above  potential.


When $a_0\delta ={\cal O} (1)$, both terms in the curly bracket of Eq.(\ref{eq:V_axion}) are comparable with each other.
In this case, it is difficult to control the potential to be flat.
On the other hand, a large $\delta $ in the factor $e^{-\delta}$ is important to make the potential flat.   
Thus, we will not discuss the case with $\delta ={\cal O}(1)$, but concentrate on the case with $\delta \gg 1$.

When $a_0\delta \ll 1$, the potential is approximated to
\begin{eqnarray}
	V 
	&\sim& A_1 \cos \theta + V_{0}- \delta(A_1-A_2) \cos(a\phi -\theta) \nonumber  \nonumber \\
	&&+ \frac{1}{2}\delta^2 (A_1-2 A_2) \cos(2a\phi-\theta)+\mathcal{O}(\delta^3),
\end{eqnarray}
where $A_1$ and $A_2$ represent the prefactors including K\"ahler potential and its derivatives,
This potential may realize the potential for  the natural inflation with modulation terms.


\section{Numerical analysis}

In this section, we investigate the poly-instanton potential (\ref{eq:Vlarge_delta}) numerically.
First, we define the slow-roll parameters,
\begin{equation}
	\varepsilon \equiv \frac{1}{2} \Bigl{(} \frac{ V_\phi}{V} \Bigr{)}^2, \qquad \eta \equiv \frac{ V_{\phi \phi}}{V},  \qquad \xi \equiv \frac{ V_\phi \cdot V_{\phi \phi \phi} }{V^2}.
\end{equation}
These parameters should be much smaller than 1 during the inflation.
We evaluate the e-folding number:
\begin{eqnarray}
	N_e(\phi_\ast) = \int^{\phi_\ast}_{\phi_{end}} d\phi \frac{V}{V_\phi} ,
\end{eqnarray}
where $\phi_{end}$ is the end-point of inflation.
The e-folding number is required to be more than 50 in order to solve the problems of the standard cosmology.
Furthermore, the Planck results constrain the power spectrum of curvature perturbation $P_\xi$, 
its spectral index $n_s$, the tensor-to-scalar ratio \cite{Planck:2013jfk,Ade:2015lrj}
\begin{eqnarray}
P_\xi = 2.20 \pm 0.10 \times 10^{-9}, \qquad 
n_s = 0.9655 \pm 0.0062, \qquad r < 0.12.
\end{eqnarray}
These can be written by use of the slow-roll parameters as,
\begin{equation}
P_\xi = \frac{V}{24 \pi^2 \varepsilon}, \qquad n_s = 1 + 2 \eta - 6\varepsilon, \qquad r = 16 \varepsilon.
\end{equation}

Table \ref{examples of inflationally potential} shows some examples of parameters and 
observables, which almost satisfy the above constrains.
The overall coefficient $A$ is decided to impose that power spectrum of curvature perturbation $P_\xi$ is $2.2\times 10^{-9}$. 
We set $a= 2\pi$ in all of the examples in the table.
The table also shows the inflation mass $m_\phi$ and the running of the spectral index $\alpha_s$, which is 
obtained as 
\begin{eqnarray}
	&&\alpha_s \equiv \frac{d n_s}{d \, ln k} \simeq -24 \varepsilon^2 + 16 \varepsilon \eta -2\xi .
\end{eqnarray}

\begin{table}[htb]
\begin{center}
  \begin{tabular}{|c|c|c|c|c|c|c|c|} \hline
      $\delta$ & $\theta$ & $A$ & $n_s$ & $r$ & $N_e$ & $\alpha_s$ & $m^2_\phi$ \\ \hline \hline
     8.0 & 4.13 & $3.29\times 10^{-16}$ & $0.953$ & $6.02\times 10^{-7}$ & $61.9$ & $-3.66\times 10^{-4}$ & $6.70\times 10^{-11}$ \\ \hline
    9.0 & 3.16 & $3.46\times 10^{-16}$ & $0.959$ & $9.30\times 10^{-7}$ & $58.6$ & $-4.66\times 10^{-4}$ & $1.27\times 10^{-10}$ \\ \hline
     10.0 & 2.11 & $2.18\times 10^{-16}$ & $0.963$ & $7.65\times 10^{-7}$ & $56.8$ & $-4.70\times 10^{-4}$ & $1.26\times 10^{-10}$ \\ \hline
    11.0 & 7.25 & $1.08\times 10^{-16}$ & $0.966$ & $4.24\times 10^{-7}$ & $61.7$ & $-3.98\times 10^{-4}$ & $8.31\times 10^{-11}$ \\ \hline
    12.0 & 6.28 & $8.89\times 10^{-17}$ & $0.966$ & $4.89\times 10^{-7}$ & $55.7$ & $-4.61\times 10^{-4}$ & $1.12\times 10^{-10}$ \\ \hline
  \end{tabular}
  \caption{Examples of parameters and observables.}
  \label{examples of inflationally potential}
\end{center}
\end{table}

\begin{figure}[tbp]
\begin{center}
\includegraphics[scale=0.6]{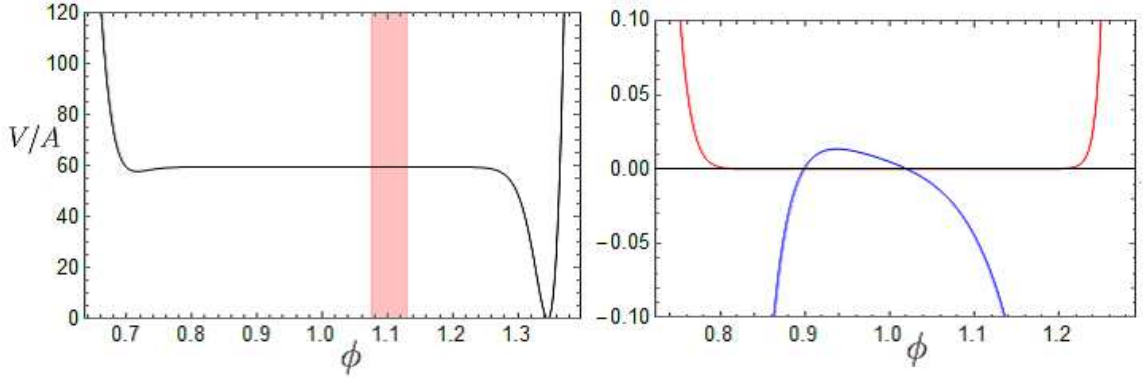}
\end{center}
\caption{The inflation potential with  $\delta=8.0$.
The red and blue curves correspond to $\varepsilon$ and $\eta$, respectively.}
\label{delta=8}
\end{figure}

\begin{figure}[htbp]
\begin{center}
\includegraphics[scale=0.6]{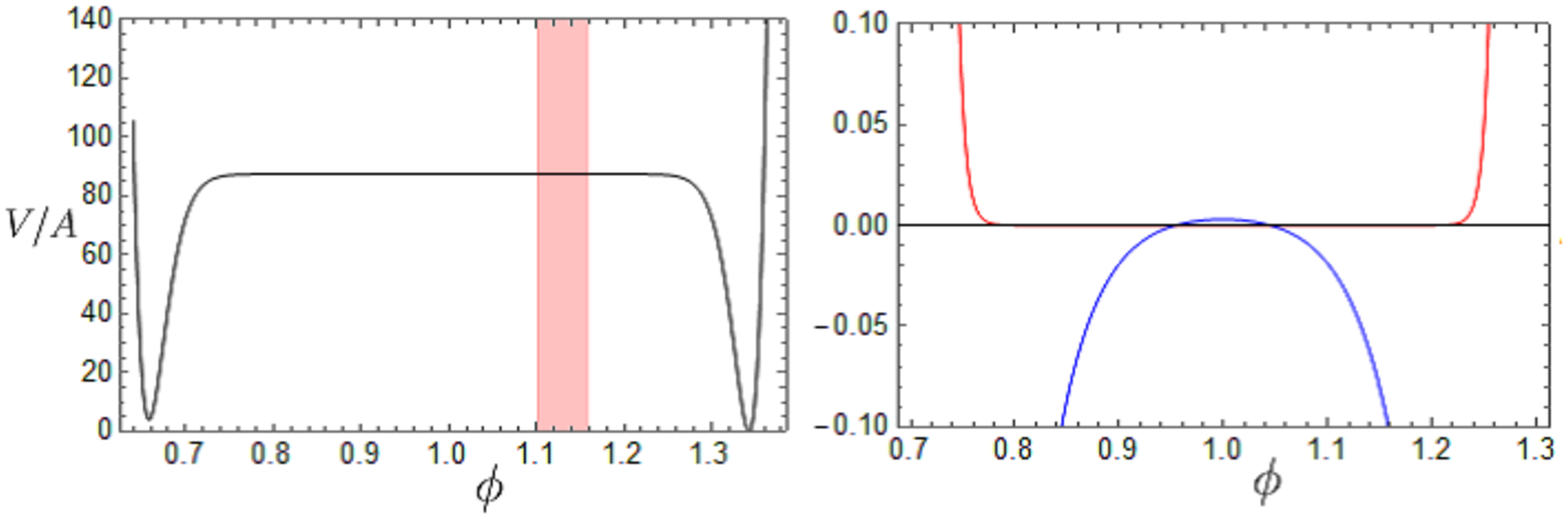}
\end{center}
\caption{The inflation potential with $\delta=9.0$.
The red and blue curves correspond to $\varepsilon$ and $\eta$, respectively.}
\label{delta=9}
\end{figure}

Figures \ref{delta=8} and \ref{delta=9} show the potential forms with $\delta = 8.0$ and $9.0$, 
which are shown in Table \ref{examples of inflationally potential}.
The left panels in the figures show the form of the potential, while the right panels show the slow-roll parameters, 
$\varepsilon$ and $\eta$.
The figures show that there is a flat region in the potential.
In order to realize  this flatness, the factor $e^{-\delta}$ is important as will be explained later.
The shaded regions correspond the field excursions during the inflation, i.e., 
$\phi_\ast$ and $\phi_{end}$.
The inflaton moves from left to right toward the potential minimum.\footnote{
Because of the potential form, there are many local minima.
the inflaton reaches one of the minima, which may be the global minimum or one of local ones.
At any rate, those minima are away the Planck scale.}

Our inflation model corresponds to the so-called small-field inflation scenario, where 
the field excursion $\Delta \phi$ during the inflation is small, i.e. $\Delta \phi < 1$.
By using the Lyth relation \cite{Lyth:1996im},
\begin{equation}\label{eq:Lyth}
r \sim 10^{-2} (\Delta \phi)^2,
\end{equation}
that implies that $r$ and $\varepsilon$ are smaller than ${\cal O}(10^{-2})$.
In order to derive $n_s = 0.9655$, we have to realize $\eta \sim -0.01$.
The right panels in Figures \ref{delta=8} and \ref{delta=9} show that 
$\eta \sim -0.01$ and $\varepsilon$ are very much suppressed.


Now, let us understand the above numerical behavior of our inflation model by 
using some approximation.
The first derivative of the potential is given by,
\begin{equation}
V_\phi = A \delta a e^{-\delta \cos a \phi}\left[  - \sin(\delta \sin (a\phi) -2 a \phi +\theta) 
+ \frac{1}{\delta} \sin(\delta \sin (a\phi) - a \phi +\theta) \right].
\end{equation}
Since we assume $\delta\gg 1$, it is approximated by
\begin{equation}
V_\phi \sim A \delta a e^{-\delta \cos a \phi} (-1) \sin(\delta \sin (a\phi) -2 a \phi +\theta).
\end{equation}
Similar to that, the second and third  derivatives can be approximated as follows,
\begin{eqnarray}
V_{\phi \phi} &\sim & A \delta^2 a^2 e^{-\delta \cos a \phi} (-1) \cos(\delta \sin (a\phi) -3 a \phi +\theta), \nonumber \\
V_{\phi \phi \phi} & \sim & A \delta^3 a^3 e^{-\delta \cos a \phi} (-1) \sin(\delta \sin (a\phi) -4 a \phi +\theta).
\end{eqnarray}
During the inflation, the potential is dominated by $A V_0$.
On the other hand, the first and second derivatives in a specific region can be much smaller because of the factor $e^{-\delta \cos a \phi }$ when $\delta$ is sufficiently large.
Thus, we can realize the flat potential without severe fine-tuning of parameters.
In addition, the ratio, $(V_\phi)^2/V_{\phi \phi}$ is suppressed by $e^{-\delta \cos a \phi} $.
Then, it is expected that $\varepsilon$ is much smaller that $\eta$ in our model.

The inflation would occur at the very flat region of the potential, i.e. $V_{\phi \phi } \approx 0$, 
and we approximate $V_{\phi \phi}$ as $V_{\phi \phi} \sim V_{\phi \phi \phi } \Delta \phi$ during inflation.
Then, we estimate the slow-roll parameter $\eta$, 
\begin{equation}
|\eta| \sim \frac{|V_{\phi \phi \phi}| \Delta \phi}{V} 
\sim \frac{A\delta^3 a^3 e^{-\delta \cos a \phi } |\sin(\delta \sin (a\phi) -4 a \phi +\theta )| \Delta \phi}{AV_0}
\sim \delta^2 a^2 \sqrt{\varepsilon} \Delta \phi .
\end{equation}
Here, the phases of sine functions in $V_\phi$ and $V_{\phi \phi \phi} $ are different from each other, 
but we estimate values of sine functions are similar.
That is, we can write  
\begin{equation}
(\Delta \phi)^2 \sim \frac{\eta^2}{\delta^4 a^4 \varepsilon}.
\end{equation}
By using the Lyth relation (\ref{eq:Lyth}),
we obtain 
\begin{equation}\label{eq:relation(r-e)}
r  \sim 10^{-2} \frac{\eta^2}{\delta^4 a^4 \varepsilon}.
\end{equation}

We have to realize $\eta \sim - 0.01$, in order to derive  $n_s = 0.9655$, as mentioned above.
Then, by use of $\eta \sim - 0.01$  and $r = 16 \varepsilon$, the relation  (\ref{eq:relation(r-e)})
leads to 
\begin{equation}
\varepsilon^2 \sim \frac{10^{-7}}{\delta^4 a^4}.
\end{equation}
For example, for $\delta = 10$, $a = 2 \pi$, we can 
estimate 
\begin{equation}
\varepsilon \sim 10^{-7}, \qquad 
r = 16 \varepsilon \sim 10^{-6} .
\end{equation}
From the relation among $P_\xi$, $\varepsilon $ and $V \approx AV_0$, we find 
\begin{equation}
V = A V_0 = 5.2\varepsilon \times 10^{-7} \sim \frac{10^{-10}}{\delta^2 a^2}.
\end{equation}
Then, the inflaton mass can be estimated as 
\begin{equation}
m_{\phi}^2 \sim V_{\phi \phi} \sim A\delta^2 a^2 e^{-\delta_{min}} \sim  {10^{-10}},
\end{equation}
where we used $V_0 \sim e^{-\delta_{min}}$.
Note that the above inflaton mass does not depend explicitly on $\delta $ and $a$.
We can etimate $m_\phi^2 \sim 10^{-10}$, i.e., $m_\phi \sim 10^{13}$ GeV.
Also, we estimate the running, 
\begin{equation}
\alpha_s =\frac{dn_s}{d \ln k} \sim  -2 \xi = -2\frac{V_\phi V_{\phi \phi \phi}}{V^2} \sim \varepsilon \eta (\delta a)^2  
\sim -10^{-3},
\end{equation}
which  does not depend explicitly on $\delta $ and $a$.
The running is negative and rather large.
A similar value has been obtained in other axion inflation models
 \cite{Kobayashi:2016vcx, Parameswaran:2016qqq}.

Finally, let us estimate the reheating temperature.
The modulus field $T_0$ would appear in the gauge kinetic function of 
the visible sector through the instanton effect,
\begin{equation}
f_{vis} = f_0+ B_{vis}e^{-b_{vis}T_0}.
\end{equation}
Through this coupling, the inflaton-axion $\phi$ can decay into 
gauge bosons, and its decay rate can be estimated by 
\begin{equation}
\Gamma \sim   \gamma^2  \left( \frac{m_\phi}{10^{13}{\rm GeV}}\right)^3~{\rm GeV},
\end{equation}
where $\gamma = B_{vis}b_{vis}e^{-b_{vis}\tau_0}$.
When this process is dominant in the decay of $\phi$, 
the reheating temperature could be estimated as 
\begin{equation}
T_{r} = \left(  \frac{\pi^2 g_\ast}{90}\right)^{-1/4} \sqrt{\Gamma M_{Pl}} \sim 
 10^{9} \gamma  \left( \frac{m_\phi}{10^{13}{\rm GeV}}\right)^{3/2}~{\rm GeV},
\end{equation}
where the effecitve degrees of freedom $g_\ast = 106.75$.
Depending on $\gamma$, the reheating temperature becomes lower.

\section{Conclusion}

We have investigated the axion potential generated by poly-instanton effects.
In superstring theory, there are $h_{1,1}$ K\"ahler moduli and they remain massless at tree level.
The $\alpha'$ correction and 1-loop correction in the K\"ahler potential can generate masses for the K\"ahler moduli, but the axions would keep flat directions.
The axions can be stabilized by non-perturbative effects and their masses could be lighter than other modulus.
Thus, the natural inflation with and without modulations would  be a good candidate for string derived inflation models.
Recently, another type of non-perturbative effects, i.e. the so-called poly-instanton effects were 
studied.
Thus, it is interesting to study the axion potential due to the poly-instanton effects and 
the possibility toward the cosmic inflation.

We have derived a new axion inflation model generated by poly-instanton effects.
The form of the poly-instanton potential is specific.
Since the derivatives of the potential are exponentially suppressed by non-perturbative effects, the potential can have a flat region 
without severe fine-tuning of parameters.
We have shown that the axion inflation is possible with our new potential, 
and that corresponds to the small-field inflation scenario.
We numerically analyze its potential and find some parameter sets realizing the inflation.
The magnitude of the poly-instanton effects is represented by the parameter $\delta$ and the large value of the $\delta$ is favored to realize the inflation. 
Most of the observables depend on the parameters, and we have shown their dependencies on $\delta a$.
For example, the tensor-to-scalar ratio is proportional to $\delta^{-4}a^{-4}$,
and it becomes smaller as $\delta a $ becomes larger. 
On the other hand, the inflaton mass is independent of $\delta$ and $a$ and is of ${\cal O}(10^{13})$ GeV.
Also, the running $\alpha_s$ is independent of $\delta$ and $a$.
Its value is negative and rather large, $\alpha_s\sim - 10^{-3}$.
A similar value has been obtained in other axion inflation models \cite{Kobayashi:2016vcx, Parameswaran:2016qqq}.
This behavior might be a generic feature in a certain class of the axion inflation models.

In this paper, we have assumed the existence of poly-instanton, but more precisely, it depends on the geometry of the compact space.
For more concrete discussion, one should define the geometry giving rise to poly-instantons.
In addition, the value of $\delta$ is significant for the poly-axion inflation, although we treat it as a free parameter.
Moreover, the poly-instanton effects in the gauge kinetic function of the visible sector would be 
important, because it affects the decay of the axion and the reheating processes.
In principle, the prefactor can be calculated as 1-loop partition function.
It may be interesting to investigate it explicitly in concrete models.
That is beyond our scope, and we would study it elsewhere.

\section*{Acknowledgments}

T. K. is supported in part by the Grant-in-Aid for Scientific Research No. 26247042 and
No. 17H05395 from the Ministry of Education, Culture, Sports, Science and Technology in
Japan.

\appendix

\section{Moduli stabilization}\label{App:example}

We give an example stabilizing the real part of a single K\"ahler modulus as well as the other moduli, 
but remaining its axionic part massless.
Suppose that there are three K\"ahler moduli fields, $T_{1,2,3}$ as well as complex structure moduli 
and the dilaton.
We assume that all of the complex structure moduli and the dilation are stabilized 
by three-form flux.
We introduce the K\"ahler potential and the superpotential as follows,
\begin{equation}
\begin{split}
W&=W_0 + A_1 e^{-a_1 T_1} + A' e^{-a' T'},\\
K&=-2 \ln \mathcal{V},\\
\mathcal{V}&= \tau_1^{3/2} - \gamma_2 \tau_2^{3/2} -\gamma_3 \tau_3^{3/2},\\
T'&= T_2 + \rho T_3.
\label{eq:w/oPoly}
\end{split}
\end{equation}
The $F$-term conditions are given as
\begin{equation}
D_{T_1} W= (K_{T_1}- a_1)A_1 e^{-a_1 T_1} +K_{T_1} (W_0 +  A' e^{-a' T'})=0,
\label{eq:F1}
\end{equation}
\begin{equation}
D_{T_2} W= (K_{T_2}- a')A' e^{-a' T'} +K_{T_2} (W_0 +  A_1 e^{-a_1 T_1})=0, 
\label{eq:F2}
\end{equation}
\begin{equation}
D_{T_3} W= (K_{T_3}- a' \rho)A' e^{-a' T'} +K_{T_3} (W_0 +  A_1 e^{-a_1 T_1})=0.
\label{eq:F3}
\end{equation}
Eqs.~(\ref{eq:F2}) and (\ref{eq:F3}) determine the ratio of $\tau_2$ and $\tau_3$, 
\begin{equation}
\begin{split}
\rho \gamma_2 \tau_2^{\frac{1}{2}} = \gamma_3 \tau_3^{\frac{1}{2}}.
\label{eq:ratio}
\end{split}
\end{equation}
Eliminating $e^{a_1T_1}$ in (\ref{eq:F1}) and (\ref{eq:F2}), we obtain
\begin{equation}
\begin{split}
\frac{a_1 a' -a_1 K_{T_2} -a' K_{T_1} }{K_{T_2}} A' e^{-a' T'} - a_1 W_0&=0.
\end{split}
\end{equation}
Similarly, eliminating $e^{-a' T'}$, we obtain the condition for $T_1$,
\begin{equation}
\frac{a_1 a' -a_1 K_{T_2} -a' K_{T_1} }{K_{T_1}} A_1 e^{-a_1 T_1}- a' W_0=0,
\end{equation}
When $a', a_1\gg1$ and $K_{T_i}$ are of the order of 1, the approximate solution is given by, 
\begin{equation}
a' T' =-\ln \left( \frac{K_{T_2} W_0}{a' A'} \right),
\end{equation}
\begin{equation}
a_1 T_1 =-\ln \left( \frac{K_{T_1} W_0}{a_1 A_1} \right),
\end{equation}
Therefore, the vacuum  expectation values of $T_1$ and $T'$ including their axionic modes are determined.
Their values can be large enough when $W_0$ is enough small.
Since the ratio of the real parts of $T_2$ and $T_3$ is determined by (\ref{eq:ratio}), there is only one massless direction, which is the imaginary part of the K\"ahler moduli which is perpendicular to $T'$ and $T_1$.
The masses of heavy modes are estimated by $W_0$. 
When $a'$ and $a_1$ is not so large, we can also tune $W_0$ such that $T_1$ and $T'$ have legitimate values.
Instead of the above superpotnetial, we could use the following superpotential:
\begin{equation}
W=W_0 + A_1 e^{-a_1 T_1} + A' e^{-a'(T_1 +Ce^{-a'' T'})},
\end{equation}
for our purpose.

Another possibility is suggested in \cite{Kobayashi:2017aeu}.
In this model, the real part of K\"ahler moduli is stabilized by 1-loop and $\alpha'$ corrections, but its imaginary part remains massless.
Similar to the above model, non-perturbative effects can stabilize its imaginary part.

More or less, we realize the models having one massless axion.
Adding to single instanton, the axion potential can be corrected by poly-instanton effects.
We can introduce relatively suppressed poly-instanton superpotential, 
$\Delta W= A \exp [b_j T_j + C e^{\sum a_i T_i}]$.
Then, the massless axion obtain a relatively light mass and the poly-instanton potential (\ref{eq:V_axion}) would appear.

\end{document}